\documentclass[aps,prb,twocolumn,superscriptaddress,floatfix]{revtex4}

\usepackage{graphicx}

\begin{document}
\newcommand{\note}[1]{{\tt{#1}}\marginpar{\hfill\rule[0mm]{3mm}{10mm}}}

\title{Quantum state preparation in semiconductor dots by adiabatic rapid passage}

\author{Yanwen Wu}\affiliation{Cavendish Laboratory, University of Cambridge, J.J. Thomson Avenue,
Cambridge CB3 0HE, United Kingdom}
\author{I.M. Piper}\affiliation{Cavendish Laboratory, University of Cambridge, J.J. Thomson Avenue,
Cambridge CB3 0HE, United Kingdom}
\author{M. Ediger}\affiliation{Cavendish Laboratory, University of Cambridge, J.J. Thomson Avenue,
Cambridge CB3 0HE, United Kingdom}
\author{P. Brereton}\affiliation{Cavendish Laboratory, University of Cambridge, J.J. Thomson Avenue,
Cambridge CB3 0HE, United Kingdom}
\author{E. R. Schmidgall}\affiliation{Cavendish Laboratory, University of Cambridge, J.J. Thomson
Avenue, Cambridge CB3 0HE, United Kingdom}
\author{P. R. Eastham}\affiliation{School of Physics, Trinity College, Dublin 2, Ireland.}
\author{M. Hugues}\affiliation{Department of Electronic and Electrical Engineering, University of
Sheffield,  Mappin Street, Sheffield S1 3JD, United Kingdom}
\author{M. Hopkinson}\affiliation{Department of Electronic and Electrical Engineering, University of
Sheffield,  Mappin Street, Sheffield S1 3JD, United Kingdom}
\author{R. T. Phillips}\affiliation{Cavendish Laboratory, University of Cambridge, J.J. Thomson
Avenue, Cambridge CB3 0HE, United Kingdom}

\date{\today}

\begin{abstract}
Preparation of a specific quantum state is a required step for a variety of proposed practical uses of
quantum dynamics. We report an experimental demonstration of optical quantum state preparation in a
semiconductor quantum dot with electrical readout, which contrasts with earlier work based on Rabi
flopping in that the method is robust with respect to variation in the optical coupling. We use
adiabatic rapid passage, which is capable of inverting single dots to a specified upper level. We
demonstrate that when the pulse power exceeds a threshold for inversion, the final state is
independent of power. This provides a new tool for preparing quantum states in semiconductor dots and
has a wide range of potential uses.
\end{abstract}

\maketitle

Preparation of a specific quantum state in a semiconductor quantum system is a required step for
quantum computation\cite{DiVincenzo95,Farhi01}, generation of single photons\cite{Michler00} and
entangled photon pairs\cite{Stevenson06}, and studies of Bose-Einstein condensation\cite{Eastham09}. A
two-level quantum system, such as that of an exciton in a single quantum dot, can be driven into a
specified state by use of a coherent interaction between the system and a tuned optical field.
Previously, the interaction used to invert a two-level system in semiconductor quantum dots has driven
the system with a resonant transform-limited light field. In this case, in the Bloch sphere
representation the Bloch vector precesses about a field vector which lies in the equatorial plane, and
so the optical pulse rotates the Bloch vector from its initial position at the south pole (ground
state) through an angle   $\theta=\pi$ to the north pole (inversion). The angle
$\theta=\int\frac{\left(\mu \cdot E \right)}{\hbar}dt$ is defined as the pulse area in a Rabi rotation
where $\mu$ is the dipole moment describing the two-level system and E(t) is the envelope of the
optical field. Coherent resonant interaction has been shown to be capable of generating several such
Rabi cycles, and permits readout of the state of the system
optically\cite{Stievater01,Kamada01,Htoon02}, or electrically by ionisation of the optical excitation
and extraction of a current\cite{Zrenner02}. The Rabi approach requires precise control over the
integrated pulse area (determined by the temporal field profile and the dipole coupling strength) to
achieve an inversion angle of $\pi$ as shown schematically in Fig. 1a.
\begin{figure}[t]
\begin{center}
\includegraphics[width=0.40\textwidth]{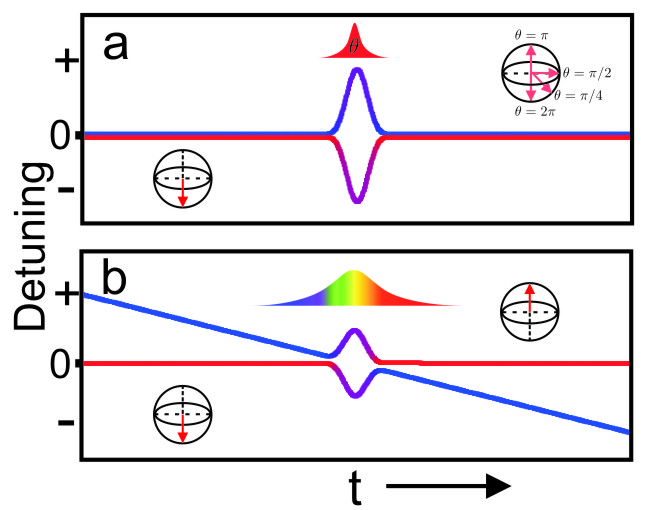}
\caption{Schematic representation of the dynamics of the two-level quantum system in time in the (a)
Rabi excitation regime with a transform-limited pulse and (b) the ARP regime  with a chirped pulse.
The curves are the eigenenergies versus time of the two levels in the rotating frame of the central
laser frequency.  The red (blue) color coding of the curves represent the population of the ground
(excited) state.  During the pulse, the admixture of colors show the mixing of the states.   As
illustrated by the Bloch spheres in (a), the final state of the system depends on the pulse
area,$\theta$, of the interacting field, whereas in (b) the system is robustly inverted independently
of the pulse area.} \label{F1}
\end{center}
\end{figure}

Here we show experimentally that state preparation is also possible by adiabatic rapid passage (ARP),
which has the advantage that it is largely unaffected by variation in the dipole coupling, which is a
normal feature of dot systems, and likewise insensitive to variation in the optical field which
typically arises from laser fluctuation or positional variation in arrays of dots\cite{Schmidgall10}.
Several theoretical proposals have recognized the potential of ARP excitation to create entanglement
between locally separated electron spins for robust two-qubit quantum
operations\cite{Hohenester06,Saikin08}, exert quantum control between two-subband quantum
wells\cite{Batista06a,Batista06b}, and generate novel Bose-Einstein condensates in
semiconductors\cite{Eastham09}.  ARP is a form of coherent interaction which effectively produces an
anticrossing of the two quantum levels involved\cite{Malinovsky01}. At an anticrossing the
wavefunction weight associated with a particular energy eigenvalue always switches from one state to
the other as the anticrossing is traversed, and ARP uses this to switch the system from the ground
state to the excited state as shown in Fig. 1b.

For ARP to operate, the quantum dynamics during the interaction with the field must not be interrupted
by random events leading to dephasing of the coherent superposition of the ground and excited
states\cite{Villas05a,Ramsay10,Villas05b}. The quantisation of electronic states in a semiconductor
quantum dot leads to an electronic level structure discrete in energy. This significantly reduces
dephasing and nonradiative recombination which proceed through phonon scattering, since the phonons
retain most features of the bulk material. Cross-gap electronic excitation remains primarily excitonic
in character, and the quantum confinement potential of the dot enhances the stability of the
multi-exciton and charged exciton species with respect to their counterparts in quantum wells or bulk
material. This leads to a complication which can be exploited as the ARP approach can be designed to
invert the quantum system to a specific final state of one of these other forms of excitation if a
suitable optical pulse is constructed\cite{Schmidgall10,Hui08}.

	In order to detect the quantum state in which the system is left by the ARP interaction we have
adopted the approach introduced by Zrenner et al.\cite{Zrenner02}, who recognised that it is possible
to exploit the separation of timescales between excitation and dephasing to read out a quantum state
electronically. In the ground state,$| 0\rangle$, there is no exciton, but the excited state, $| X
\rangle$, corresponds to an electron in the conduction band and a hole in the valence band, bound by
the mutual Coulomb interaction as an exciton. The resonant optical excitation takes the system from
the ground state to an arbitrary coherent superposition with the upper state: $c_0| 0 \rangle +c_X| X
\rangle$. The dot is embedded in a biased Schottky diode structure (Fig. 2), and the applied electric
field leads to ionisation of the excited state on a timescale longer on average than the excitation
time. Clearly, the probability of charge flow depends on the relative amplitude in state $| X
\rangle$; when the system is entirely inverted and $| c_X |^2=1$, the current flow in the external
circuit is one electron per excitation cycle. For a repetition of the incident laser pulse at a rate f
= 76 MHz, the peak current expected in this simple picture is just ef where e is the electron charge;
in our experiment ef=12.2 pA.

\begin{figure}[t]
\begin{center}
\includegraphics[width=0.50\textwidth]{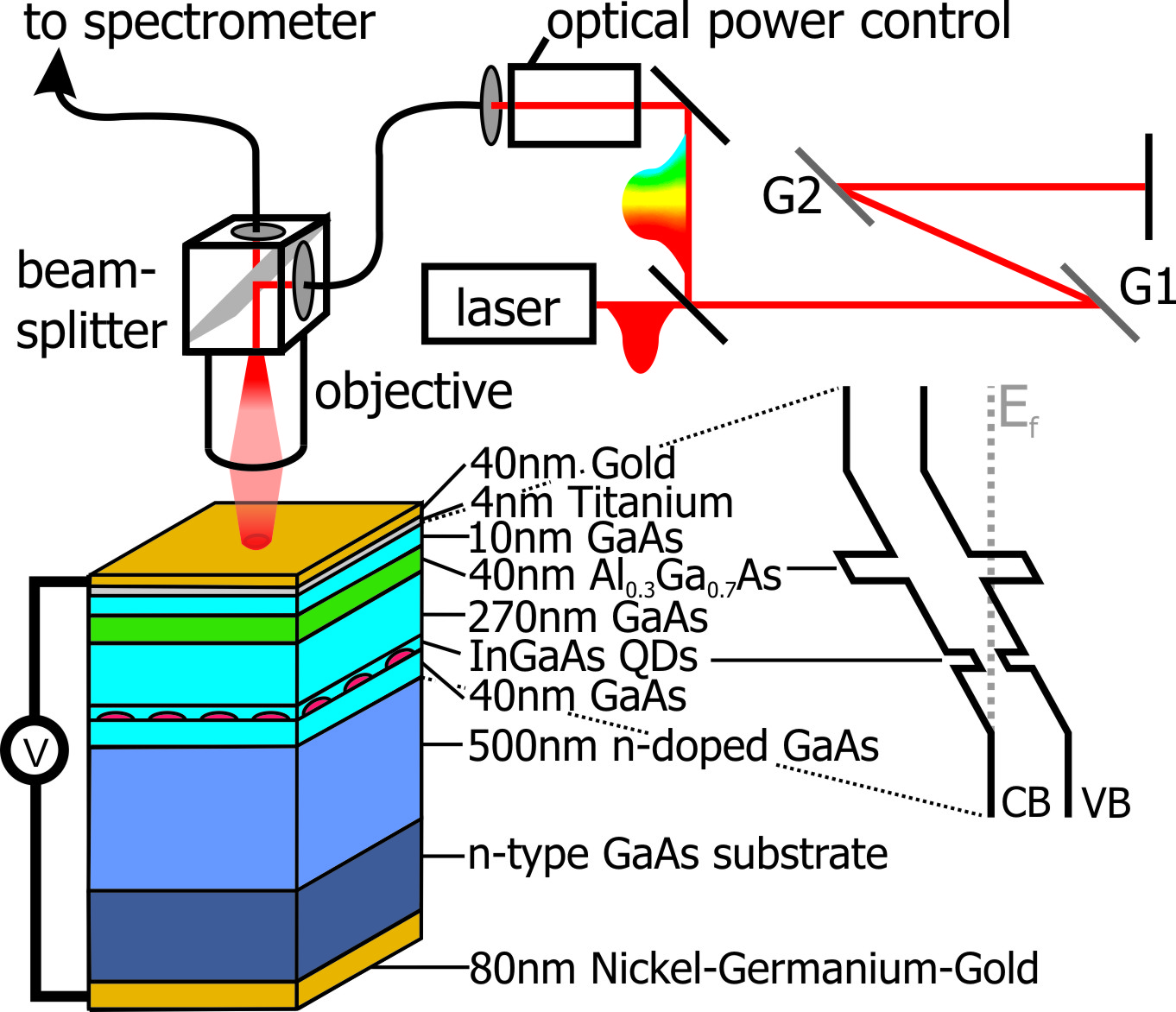}
\caption{Schematic diagram of the sample structure and experimental setup.  The linear chirp on the
optical pulses is generated by a parallel grating pair with 1200 grooves/mm.  The value of the chirp
can be changed by varying the distance between the gratings.  The laser light is focused by the
objective down to ~1$\mu$m over a 200 nm aperture in a gold mask.  The sample consists of a single
layer of InGaAs quantum dots embedded in a Schottky diode structure.  The band diagram of the diode
shows the relative positions of each layer in the growth direction and in energy.} \label{F2}
\end{center}
\end{figure}

	We have selected a single InGaAs dot formed by Stranski-Krastanow growth, observed through a 200
nm diameter aperture fabricated in a Ti/Au Schottky contact by electron-beam lithography, as
illustrated schematically in Fig. 2.  Within the structure the dot layer is separated from the heavily
n-doped back layer by 40nm of GaAs which acts as a tunnelling barrier. The position of the dots with
respect to the Fermi level can be changed by varying the bias applied to the top Schottky contact. The
aperture is illuminated by a confocal system and light collected for spectroscopy. Illumination is at
a photon energy higher than the gap for excitation of photoluminescence (PL) or by pulses from a
mode-locked Ti:Sapphire laser for the resonant pulse experiments. The selection of the laser
wavelength for the pulsed experiments is made by first conducting photoluminescence mapping of the
transitions corresponding to this dot, as shown in Fig. 3a.

\begin{figure}[t]
\begin{center}
\includegraphics[width=0.50\textwidth]{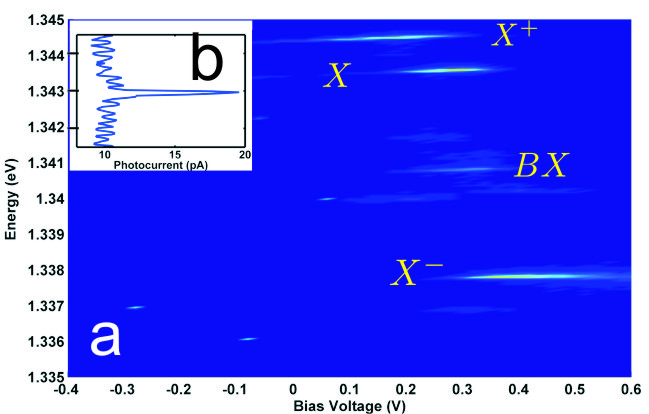}
\caption{A PL map of the emission from various excitonic  states at different bias voltages.  $X^+$ is
the positively-charged exciton, X is the neutral exciton, BX is the neutral biexciton, and $X^-$ is
the negatively-charged exciton.  (b) Photocurrent versus the wavelength of a continuous-wave laser at
-1 V bias voltage.} \label{F3}
\end{center}
\end{figure}
Note that the dot can be switched from the negatively-charged exciton ($X^-$ , at about 1.338 eV) to
the neutral exciton (X, at about 1.3435 eV) and to the positively-charged exciton (X+, at about 1.3445
eV) by varying the bias on the device, demonstrating the charge injection by tunnelling. Also present
is the recombination of the first exciton pair in the biexciton state (BX) which is emitted at about
1.341 eV corresponding to a biexciton binding energy of around 3 meV. In order to read the quantum
state of the dot by means of the ionisation current the bias has to be chosen to suppress
photoluminescence as the main recombination channel. We have chosen to operate the device at a bias of
-1 V, which suppresses the PL signal but does not produce too short a tunnelling time\cite{Findeis01}.
Fig. 3b shows the photocurrent at that bias as a function of tunable continuous-wave laser wavelength
incident on the structure; the photocurrent peak at 1.343 eV corresponds to resonant excitation of X,
with the transition energy modified slightly by the applied field.

\begin{figure}[t]
\begin{center}
\includegraphics[width=0.50\textwidth]{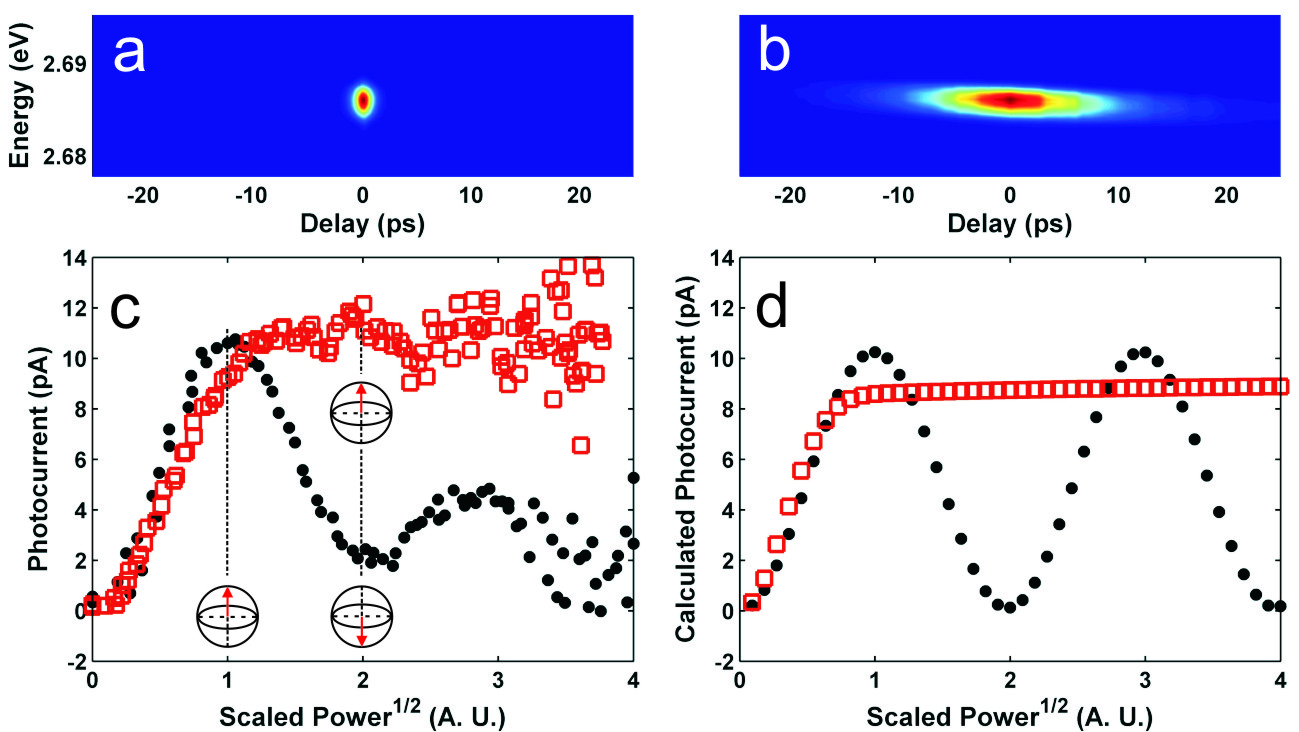}
\caption{(a) An up-converted spectrally resolved autocorrelation scan of the transform-limited pulse
of 2 ps at FWHM and (b) an up-converted spectrally resolved crosscorrelation scan of the chirped pulse
with a transform-limited pulse of 2 ps FWHM.  (c) Electrical readout signals of Rabi oscillation
(solid circles) excited by the pulse characterized in (d) and ARP (open squares) excited by the pulse
characterized in (b).  The curves are the difference between the on-resonant (1.343 eV) photocurrent
and the corresponding off-resonant (1.341 eV) background photocurrent. The small Bloch spheres
indicate the position of the Bloch vector following excitation at the given power in each case.  (d)
Simulation of the corresponding electrical readout signals in (c) using the same optical pulse
characterization and sample properties as the experiment assuming a recombination time of 1 ns and
tunnelling time of 300 ps.} \label{F4}
\end{center}
\end{figure}

 Under pulsed excitation with a transform-limited pulse with zero temporal chirp ($\alpha =
0$ ps$^{-2}$; see definition below) of 2 ps full width at half maximum (FWHM) (Fig. 4a), tuned to
coincide with the photocurrent peak, the system clearly exhibits Rabi oscillation as the pulse area
(proportional to [incident power]$^{\frac{1}{2}}$) is increased, as shown in Fig. 4a by the black
symbols. The first peak of the Rabi flop corresponds to the exciton being left in the excited state
after the pulse has passed (Bloch vector toward the north pole), whereas the first trough corresponds
to a Rabi rotation through the exciton and back to the ground state (Bloch vector toward the south
pole).  These data in Fig. 4c represent the difference between the current drawn when the spectrum of
the pulse overlaps the transition and when it is detuned from it. The presence of a background current
has been reported in  previous work on electrical readout of this form\cite{Zrenner02,Ramsay10}, and
is associated with weak absorption processes anywhere within the depletion region of the device. When
the same optical pulse is chirped by the grating pair (G1 and G2 in Fig. 2) to give a temporal chirp
of $\alpha = 0.089$ ps$^{-2}$, and a chirped pulse width of 15 ps at FWHM (Fig. 4b),  the result is a
clear signature of adiabatic rapid passage, as shown in Fig. 4a by the red symbols, where the current
rises initially as the pulse power is increased, and then stabilises at a value corresponding to
inversion, with no further change as the pulse area is increased. Thus at an equivalent pulse power to
that which left the system back in the ground state, with ARP the dot is left inverted. With the
bandwidth used here (0.3 meV) there is no contamination of the inversion from the biexciton
transition.

We can compare the measured readout of the quantum state of the system with a calculation of the
current using the methods described by Villas-Bôas et al.\cite{Villas05a}, and Schmidgall et
al.\cite{Schmidgall10}.  This system can be described by the Hamiltonian
\begin{equation}
H=E_X| X \rangle \langle X |-\frac{\mu_X}{2}\left(E(t)| 0 \rangle \langle X | + E^*(t)| X \rangle
\langle0 | \right)
\end{equation}.
The dipole coupling is specified by $\mu_X$ and the optical field by E(t). Under a unitary
transformation
\begin{equation}
U(t)=\exp{\left[0| 0 \rangle\langle0 | + i \omega(t)t| X \rangle \langle X |\right]}
\end{equation}
this yields a picture appropriate to interaction driven by an applied field whose frequency varies in
time as $\omega(t)$. The simplest variation is linear in time, with $\omega(t)=\omega_0+\frac{\alpha
t}{2}$, where $\alpha$ is the linear temporal chirp. This is the form of optical field produced by the
grating pair in our setup shown in Fig. 2.  The Hamiltonian in the rotating frame of the central
frequency of the laser, $\omega_0$, is
\begin{equation}
H_{\mathrm{eff}}=\left(
\begin{array}
{cc}
0& -\frac{1}{2}\mu_XE_0(t)\\
-\frac{1}{2}\mu_XE_0(t)&\Delta_X-2 \hbar \dot{\omega}(t)\\
\end{array}
\right)
\end{equation}

where the detuning of  $\omega_0$ from the transition frequency of the exciton is $\Delta_X$.  In the
experiment, $\Delta_X$ is zero.  In this picture, the condition for adiabatic transfer (extensively
explored previously\cite{Malinovsky01,Malinovskaya09}) is that the effective Rabi frequency
$\Omega(t)=\sqrt{|\Omega_0(t)|^2+\delta(t)^2}$ satisfies $\frac{\dot{\Omega}_0}{\Omega^2}\ll 1$, and
$\frac{\dot{\delta}}{\Omega^2}\ll 1$, where $\hbar \delta=\Delta_X - \hbar\dot{\omega}t$ and
$\Omega_0(t)=\mu_X E_0(t)/\hbar$.   In this adiabatic regime, the Bloch vector follows the field
vector as it rotates at a rate of $\dot{\delta}$ from one polar extreme to another during the pulse
while precessing rapidly at $\Delta$ around the field vector within a small solid angle.

To calculate the current drawn from the dot in the presence of both dephasing and ionisation of the
exciton, the Hamiltonian model of the individual system is used to evaluate the time evolution of the
density matrix; the current is calculated from the scattering term corresponding to taking state $|X
\rangle$ to state $|0\rangle$ by ionisation, using the Lindblad form\cite{Lindblad76}  of the
scattering terms as described by Schmidgall et al.\cite{Schmidgall10}. This term is integrated
throughout the pulse interaction. Note that this model does not incorporate terms intended to explain
the reduction in contrast of the Rabi oscillation usually observed for high values of pulse area in
Rabi flopping\cite{Ramsay10,Villas05b,Brandi05,Vagov07,Wang05}. For realistic values of dephasing and
tunnelling parameters, the model generates the curves shown in Fig. 4d, which confirm that the
measured signals correspond to ARP.

Our results demonstrate the possibility of quantum state inversion in a system measured by electrical
readout, robust with respect to variation in the details of the strength of the optical interaction.
This opens the possibility of using adiabatic rapid passage in a range of contexts, including
inversion of systems with level structures which can lead to deterministic single photon emission, or
entangled photon pair emission. Adiabatic rapid passage requires only a chirp of the relative
frequency offset of the optical field and transition frequency, so can be driven by a suitably rapid
electrical pulse sweeping the transition. The insensitivity to the details of the interaction can be
expected to provide access for the first time to physics associated with injection of tailored
inversion profiles, such as complex microcavity electrodynamics.

\begin{acknowledgements}
This work is supported by EPSRC grant EP/F040075/1. YW is grateful for a Herchel Smith Fellowship. We
would like to thank Prof. Dr. Artur Zrenner for his valuable advice on the fabrication of a
low-leakage current Schottky device for electrical readout, Dr. Paul Eastham for helpful discussions
and Dr. Geb Jones and Jonathan Griffiths for invaluable assistance with electron beam lithography.
\end{acknowledgements}

\end{document}